\documentclass[a4paper,11pt]{article}
\pdfoutput=1 

\usepackage{jcappub} 
\usepackage{subcaption}
\usepackage[T1]{fontenc} 
\usepackage{sidecap}
\usepackage{subcaption}
\usepackage{booktabs}
\usepackage{array}
\usepackage{cleveref}
\usepackage{xcolor}
\usepackage{soul}

\title{Observational Constraints on\\ Warm Natural Inflation}

\author[a,1]{Gabriele Montefalcone,\note{Corresponding author.}}
\author[a]{Vikas Aragam,}
\author[b,c]{Luca Visinelli}
\author[a,d,e]{and Katherine Freese}

\affiliation[a]{Texas Center for Cosmology and Astroparticle Physics, Weinberg Institute for Theoretical Physics, Department of Physics, University of Texas, Austin, Texas 78751, USA}
\affiliation[b]{Tsung-Dao Lee Institute (TDLI),
520 Shengrong Road, 201210 Shanghai, P.\ R.\ China}
\affiliation[c]{School of Physics and Astronomy, Shanghai Jiao Tong University,
800 Dongchuan Road, 200240 Shanghai, P.\ R.\ China}
\affiliation[d]{The Oskar Klein Centre, Department of Physics, Stockholm University, AlbaNova, SE-10691 Stockholm, Sweden}
\affiliation[e]{Nordic Institute for Theoretical Physics (NORDITA), 106 91 Stockholm, Sweden}

\emailAdd{montefalcone@utexas.edu}
\emailAdd{aragam@utexas.edu}
\emailAdd{luca.visinelli@sjtu.edu.cn}
\emailAdd{ktfreese@utexas.edu}

\abstract{Warm natural inflation is studied for the case of the original cosine potential.
The radiation bath during inflation induces a dissipation (friction) rate in the equation of motion for the inflaton field, which can potentially reduce the field excursion needed for an observationally viable period of inflation.
We examine if the dissipation thus provides a mechanism to avoid the large decay constant $f \gtrsim M_{\mathrm{pl}}$ of cold cosine natural inflation. Whereas temperature independent dissipation has previously been shown to alleviate the need for a trans-Planckian decay constant $f$, we illustrate here the difficulties of accommodating a significantly sub-Planckian decay constant ($f<10^{-1}M_{\mathrm{pl}}$) for the case of a temperature dependent dissipation rate in the form $\Gamma \propto T^c$, with $c=\{1,3\}$. Such dissipation rates represent physically well-motivated constructions in the literature. For each model, we map its location in the $r$-$n_s$ plane 
and compare with cosmic microwave background (CMB) data. For $c=1 \, (c=3)$, we find that agreement with CMB data requires that dissipation be in the weak (moderate) regime and that the minimum allowed value of the decay constant in the potential is $f_{\rm min} = 0.3 \, (0.8)\,M_{\mathrm{pl}}$, respectively.
}

\preprint{UTWI-24-2022, NORDITA 2022-089}
\begin{document}
\maketitle
\flushbottom

\section{Introduction}
\label{sec: introduction}

Inflation~\cite{Guth:1980zm, Linde:1981mu, Albrecht:1982wi, Kazanas:1980tx, Starobinsky:1980te, Sato:1980yn, Mukhanov:1981xt, Linde:1983gd, Mukhanov:1990me} is currently the most convincing mechanism to address the horizon, flatness, and monopole problems in the standard Big-Bang cosmology. The accelerated expansion rate during inflation ensures that when the process ends, the Universe is sufficiently flat, homogeneous and isotropic at the largest observable scales.

In addition to addressing these problems, inflation provides a mechanism for generating the density fluctuations that later evolve into the large structures observed in the cosmic web. The pattern in the spectrum of fluctuations should be consistent with the angular power spectrum observed in the cosmic microwave background (CMB) by an array of experiments, most recently the {\it Planck} satellite~\cite{Planck:2018nkj}. In standard inflationary models, these fluctuations are quantum-mechanical in origin and adiabatic. Field-theoretic constructions attribute this quantum nature to the fluctuations of a scalar field responsible for driving the expansion rate during inflation, namely the {\it inflaton} field.

Conventional models of inflation involve a single scalar field slowly rolling down a nearly flat potential, inducing a quasi-de Sitter phase. Although the inflaton is often taken to be only (minimally) coupled to gravity, introducing couplings to other early universe sectors can relax various restrictions normally imposed in inflationary models. A well-established alternative framework to conventional inflation is \textit{warm inflation}, in which the inflaton is thermally coupled to a bath of radiation~\cite{Berera:1995ie, Berera:1996fm}. Fluctuations in warm inflation are predominantly thermal in origin, with quantum fluctuations being subdominant in the limit of a large dissipation rate between the two sectors. Additionally, the inflaton continually sources the production of radiation, which alleviates the need for a separate reheating phase at the end of inflation. In this paper, we study the model of warm natural inflation.

The basic idea behind natural inflation~\cite{Freese:1990rb} is to use an axion as the inflaton to provide a ``natural'' explanation of the flat potential required for inflation. As shown by ref.~\cite{Adams:1990pn} in the context of cold inflation models and generalized by ref.~\cite{Montefalcone:2022owy} to warm inflation, potentials must be extremely flat for rolling models of inflation. The potential must be wide enough to sustain a sufficient number of $e$-folds of inflation (typically 60), while its height is restricted in scale in order to not overproduce density perturbations from quantum fluctuations. This combination is typically difficult to achieve in particle physics models, in which both the height and width of the potential are set by the same scale. For example, for $V(\phi) = \lambda \phi^4$, inflation requires $\lambda \sim 10^{-12}$ whereas loop corrections due to the coupling of $\phi$ to any other field typically drive $\lambda \sim \mathcal{O}(1)$. This ``fine-tuning'' problem is resolved in natural inflation, in which the inflaton potential is protected against these loop corrections by a shift symmetry, e.g.\ the inflaton may be a pseudo Nambu-Goldstone boson. The original model considered was obtained by directly mimicking the physics of the QCD axion at higher scales; the idea of using an axion as the inflaton of course generalizes to other models. 

The simplest version of natural inflation to consider is still the original model, in which the potential takes a cosine form as considered in this paper; see eq.~\eqref{eq:3.1} below. The downside of the original cosine potential is that the latest CMB data seem to require the axion decay constant $f$ to lie close to or above the Planck scale \cite{Freese:2014nla,Stein:2021uge}, and many physicists are concerned about trans-Planckian effects on the potential. In addition, such large values of $f$ cannot easily be accommodated in string theory~\cite{Banks:2003sx}. Hence, variations on the original cosine potential have been studied, such as aligned natural inflation~\cite{Kim:2004rp} and axion monodromy~\cite{Silverstein:2008sg}.

In this paper, we consider warm natural inflation for the case of the original cosine potential~\cite{Mishra:2011vh, Visinelli:2011jy}. One goal of the paper is to see whether or not the trans-Planckian values of the decay constant $f$ can be avoided in the presence of the radiation bath of warm inflation. The radiation bath produced while the inflaton is rolling down its potential effectively acts like a friction term. Hence, given the same number of $e$-folds,
the field excursion $\Delta \phi$ can be reduced compared to the cold inflation case,
i.e.\ the width of the potential could be smaller while still obtaining sufficient inflation.

One might then naively think that sub-Planckian values of $f$ could easily be accommodated. In the earliest studies of warm natural inflation~\cite{Visinelli:2011jy}, this conclusion was true. However, those studies assumed a constant rate of dissipation throughout the epoch of inflation. Here, on the other hand, we consider that the dissipation rate has a temperature dependence $\Gamma \propto T^c$ specifically with $c=\{1,3\}$, in accordance with physically motivated axion-like interactions with bosonic and fermionic fields~\cite{Visinelli:2014qla, Visinelli:2016rhn, Berghaus:2019whh, Berghaus:2020ekh, DeRocco:2021rzv}. In order to compare these models with CMB data, we will compute the CMB observables for the models (spectral index $n_s$ and tensor to scalar ratio $r$) and then show the resulting plots in the $r$-$n_s$ plane; see figure~\ref{fig:1}. We find that warm natural inflation is consistent with CMB observations for a marginally sub-Planckian decay constant $f$. This result is valid for both cases $c=1$ and $c=3$.

Here we mention previous related work.
For a recent model of warm natural inflation with a non-minimal coupling to gravity and a linear temperature-dependent dissipation rate see ref.~\cite{AlHallak:2022haa}; our work is the first study for $c=1$ dissipation with Einstein gravity. The cubic case ($c=3$) was studied in ref.~\cite{Reyimuaji:2020bkm} and recently in ref.~\cite{Correa:2022ngq}. We disagree with the results of ref.~\cite{Reyimuaji:2020bkm}, who claimed more stringent bounds on $f$, but we believe there were some errors in the calculation.\footnote{The authors of ref.~\cite{Reyimuaji:2020bkm} claim that consistency with \textit{Planck} 2018 requires $f\gtrsim 3 M_{\mathrm{pl}}$. This is however due to the fact that they do not explicitly fix the magnitude of the primordial power spectrum to match the CMB observations but instead set a priori the scale of the inflaton field to $10^{16}\,\mathrm{GeV}$.
Additionally, they also mistakenly take $\beta_w\neq 0$, see eq.~\eqref{eq:SRp3}, which we instead set to $0$ since $\Gamma$ in this work has no explicit $\phi$ dependence.} The authors of ref.~\cite{Correa:2022ngq} found that warm natural inflation with $c=3$ is consistent with \textit{Planck} 2018 for a marginally sub-Planckian $f$ (we agree with their results) and that this also provides the perfect conditions for the production of primordial black holes in the mass range where it can account for all of the the dark matter content of the universe. In this paper, we perform a more extensive study for the case of $c=3$ by analyzing the parameter space compatible with the CMB observational constraints, as a function of the decay constant $f$ and the strength of the dissipation; and by imposing various constraints for successful inflation. We also compare the derived constraints with the thermal field theory requirements on the physically-well motivated axion-like interaction terms that produce a linear and cubic dissipation rate.\footnote{See also ref.\cite{Klose:2022rxh} for a recent study of warm natural inflation with a dissipation rate motivated by updated lattice calculations in ref.\cite{Laine:2022ytc}, which in its closed form is not a power-law of the temperature.}

We begin in section~\ref{sec:2} with the basics of the warm inflation scenario as well as the calculation of the perturbation spectra, with particular emphasis to the enhancement in the amplitude of the power spectrum caused by thermal effects, see eq.~\eqref{eq:PS}. Then in section~\ref{sec:3} we turn to the warm natural inflation scenario. We find the bounds on parameters to require the existence of a broad slow-roll regime. We calculate the number of $e$-folds of inflation, and then the CMB observables. In section~\ref{sec:results} we present our results for the cases of a linear and cubic temperature-dependent dissipation and illustrate in detail the difficulties of accommodating a significantly sub-Planckian decay constant $f$. We conclude with a summary in section~\ref{sec:5}.

\section{The warm inflation scenario
\label{sec:2}}
\subsection{Basic framework}

In the warm inflation scenario, the inflaton field substantially converts into radiation already during the inflationary period. The dissipation mechanism is parameterized by the introduction of a non-negligible dissipation rate $\Gamma$ in the dynamics of the inflaton field: 
\begin{equation}
    \Ddot{\phi}+(3H+\Gamma)\dot{\phi}+V_{,\phi}=0\,,
    \label{eq:2.1}
\end{equation}
where a dot denotes the derivative with respect to cosmic time and $V_{,\phi}=\partial V/\partial \phi$. The criterion for warm inflation to occur is that the thermal fluctuations dominate over the quantum fluctuations. This simply amounts to $H<T$~\cite{Berera:1995ie}.

In the following, we assume that the radiation thermalizes on a time scale much shorter than $1/\Gamma$~\cite{Berera:1995wh,Berera:1995ie}, so that the energy density of radiation can be taken to be:
\begin{equation}
    \rho_R(T)=\alpha_1 T^4, \quad \text{with } \quad \alpha_1=\frac{\pi^2}{30}g_{*}(T), \label{eq:2.2}
\end{equation}
where $g_{*}(T)$ is the number of relativistic degrees of freedom of radiation at temperature $T$. Although we do not specify $g_{*}(T)$ in the equations, we adopt the number of relativistic degrees of freedom in the minimal supersymmetric Standard Model $g_{*}(T) = 228.75$ when presenting the results in section~\ref{sec:results}.

From energy conservation, we obtain the evolution of the radiation energy density,
\begin{equation}
    \Dot{\rho}_R+4H\rho_R=\Gamma \Dot{\phi}^2\,,  \label{eq:2.3}
\end{equation}
with the term on the right-hand side of eq.~\eqref{eq:2.3} representing the energy transferred from the inflaton to the radiation bath.
Finally, the Friedmann equation for the background evolution reads:
\begin{equation}
    H^2=\frac{1}{3M_{\mathrm{pl}}^2}\left[V(\phi)+\frac{1}{2}\Dot{\phi}^2+\rho_R\right]\,,  \label{eq:2.4}
\end{equation}
where $M_{\mathrm{pl}} = 1/\sqrt{8 \pi G} \, \approx 2.435\times 10^{18}\,\mathrm{GeV}$ is the reduced Planck mass.

Inflation is realized when the Hubble expansion rate $H$ is approximately constant. This is achieved when the potential $V(\phi)$ is approximately flat and the potential energy dominates over all other forms of energy. During this period, known as the slow-roll regime of the inflaton field, higher order derivatives in eqs.~\eqref{eq:2.1} and~\eqref{eq:2.3} can be neglected,
\begin{equation}
    \ddot{\phi} \ll H \dot{\phi}, \quad \text { and } \quad \dot{\rho}_{R} \ll H \rho_{R}\,. \label{eq:2.5}
\end{equation}
As a result, in this regime the background evolution and the equation of motion for the inflaton and the radiation bath respectively read:
\begin{align}
    H^2&\simeq \frac{V}{3M_{\mathrm{pl}}^2}\,, \label{eq:SR1} \\
    \Dot{\phi}&\simeq -\frac{V_{,\phi}}{3H(1+Q)}\,, \label{eq:SR2}\\
    \rho_r&\simeq \frac{3Q \Dot{\phi}^2}{4}\,,\label{eq:SR3}
\end{align}
where the dimensionless ratio $Q$ measures the effectiveness at which the inflaton converts into radiation and it is defined as
\begin{equation}
    Q \equiv \frac{\Gamma}{3 H}\,.
\end{equation}
For $Q\lesssim 1$ we achieve a weak regime of warm inflation. For the purpose of this paper, we define a moderately dissipative regime $1\leq Q\leq 10$, and the strongly dissipative regime with $Q \gg 10$.\footnote{Note, that in the literature the definition $Q \gg 1$ is usually adopted for the strong regime.} We show that warm natural inflation is consistent with data for the weak and moderately dissipative regimes.

We can parameterize the slow-roll regime in warm inflation (subscript ${w}$) via a set of three parameters $\epsilon_{w}$, $\eta_{w}$ and $\beta_{w}$, defined by:
\begin{align}
\epsilon_{w} & \equiv \frac{\epsilon_{V}}{1+Q}=\frac{M_{\mathrm{pl}}^{2}}{2(1+Q)}\left(\frac{V_{, \phi}}{V}\right)^{2}\,, \label{eq:SRp1} \\
\eta_{w} & \equiv \frac{\eta_{V}}{1+Q}=\frac{M_{\mathrm{pl}}^{2}}{(1+Q)}\left(\frac{V_{, \phi \phi}}{V}\right)\,, \label{eq:SRp2} \\
\beta_{w} & \equiv \frac{M_{\mathrm{pl}}^{2}}{(1+Q)}\left(\frac{\Gamma_{, \phi} V_{, \phi}}{\Gamma V}\right) \label{eq:SRp3}\,.
\end{align}
In warm inflation, the slow-roll regime is achieved whenever $\epsilon_w \ll 1$, $|\eta_w| \ll 1$ and $|\beta_w| \ll 1$ are all satisfied~\cite{Taylor:2000ze, Hall:2003zp, Bastero-Gil:2004oun, Moss:2008yb}. Eqs.~\eqref{eq:SRp1},~\eqref{eq:SRp2} and~\eqref{eq:SRp3} can be thought of as a generalization of the slow-roll conditions obtained in cold inflation, which take into account dissipation through the parameter $Q$. For $Q\gg 1$, the slow-roll conditions are substantially relaxed and can in principle be satisfied by scalar field potentials that would otherwise violate the standard slow-roll conditions in the cold inflation scenario~\cite{Motaharfar:2018zyb,Das:2020xmh,Kamali:2021ugx,Santos:2022exm}. 

For $\Gamma=\Gamma(\phi, T)$ and $V_{,T} \equiv \partial V/\partial T =0$, we can differentiate eqs.~\eqref{eq:SR1},~\eqref{eq:SR2} and~\eqref{eq:SR3} to obtain the following identities:
\begin{align}
    \frac{{\rm d}\ln H}{{\rm d}N_e}&=-\epsilon_w\,, \label{eq:d1} \\
    \frac{{\rm d}\ln T}{{\rm d}N_e}&=\frac{1}{(4-c)+Q(4+c)}\Big[(3+Q)\epsilon_w-(1+Q)2\eta_w- (1-Q)\beta_w\Big]\,, \label{eq:d2} \\
    \frac{{\rm d}\ln \Dot{\phi}}{{\rm d}N_e}&=-\eta_w+\epsilon_w+\frac{2Q}{(4-c)+Q(4+c)}\Big[c\eta_w- (2+c)\epsilon_w+ 2\beta_w\Big]\,, \label{eq:d3} \\
    \frac{{\rm d}\ln Q}{{\rm d}N_e}&= -\frac{2(1+Q)}{(4-c)+Q(4+c)}\Big[c\eta_w- (2+c)\epsilon_w + 2\beta_w\Big]\,, \label{eq:d4} 
\end{align}
where ${\rm d}N_e=H{\rm d}t$ is the differential increment in the number of $e$-folds, discussed in more detail around eq.~\eqref{eq:3.9} below, and the coefficient $c$ describes the dependence of the dissipation rate on temperature as
\begin{equation}
    \label{def:c}
    c \equiv \frac{T}{\Gamma}\Gamma_{,T}\,.
\end{equation}
Demanding that the radiation bath is generated at a faster rate than redshift so that a successful reheating is achieved during inflation leads to the bound $c < 4$~\cite{Bastero-Gil:2011rva}.

\subsection{Perturbation spectra}

The presence of a radiation bath and a dissipation rate not only alters the background dynamics of the inflaton field but also its perturbations. Specifically, the addition of these thermal effects can have a significant impact on the primordial power spectrum, which in warm inflation takes the general form~\cite{Hall:2003zp, Graham:2009bf, Bastero-Gil:2011rva, Ramos:2013nsa}:
\begin{equation}
    \label{eq:PS}
    \Delta_{\mathcal{R}}^{2}=\left(\frac{H^{2}}{2 \pi \dot{\phi}}\right)^{2}\left[1+2 n_{\mathrm{BE}}+\frac{2 \sqrt{3} \pi Q}{\sqrt{3+4 \pi Q}} \left(\frac{T}{H}\right)\right] G(Q)\,,
\end{equation}
where the quantities are all evaluated at horizon crossing. Here, $n_{\mathrm{BE}} =1/\left[\exp(H/T)-1\right]$ is the Bose-Einstein distribution function which reflects the inflaton statistical distribution due to the presence of the radiation bath. As expected, in the limit $Q \to 0$ and $T \to 0$, eq.~\eqref{eq:PS} reproduces the standard cold inflation result. When $T>H$, the second and third terms in the square bracket dominate as they account for the thermal contributions to the inflaton fluctuations.

The expression in eq.~\eqref{eq:PS} contains a function $G(Q)$ that accounts for the coupling of the inflaton and radiation fluctuations due to a temperature-dependent dissipation rate, and which can only be determined numerically by solving the full set of perturbation equations~\cite{Graham:2009bf, Bastero-Gil:2011rva}. For a constant value of $\Gamma$, corresponding to $c=0$ in eq.~\eqref{def:c}, fluctuations do not further affect the spectrum and $G(Q) = 1$, while $G(Q)$ is generally larger (smaller) than one for $c >0$ ($c<0$). Numerical fits for $G(Q)$ in terms of a ratio of two polynomials in $Q$ have been reported in Refs.~\cite{Bastero-Gil:2016qru, Benetti:2016jhf} for the cases $c = 1$ and $c=3$.

The result in eq.~\eqref{eq:PS} from the theory of perturbations at horizon crossing is assessed against the amplitude of the scalar power spectrum measured by the CMB radiation data at a pivotal scale $k_*$. The \textit{Planck} Collaboration reports the result from TT,TE,EE+lowE+lensing datasets at 68\% confidence level (CL) as $\ln(10^{10}A_s) = 3.044\pm 0.014$~\cite{Planck:2018vyg}, where $A_s = \Delta_{\mathcal{R}}^2(k_{*})$ is the power spectrum at the scale $k_{*} = 0.05\,\mathrm{Mpc}^{-1}$, corresponding to $\Delta_{\mathcal{R}}^2(k_{*}) \simeq 2.1 \times 10^{-9}$.

Given the scalar curvature power spectrum, the spectral tilt $n_s$ and the tensor to-scalar ratio $r$ are defined as in the cold inflation case:
\begin{align}
    n_s-1& \equiv \frac{{\rm d}\ln \Delta^2_{\mathcal{R}}}{{\rm d}\ln k}\simeq \frac{{\rm d}\ln \Delta^2_{\mathcal{R}}}{{\rm d}N_e},\label{eq:ns1} \\
    r& \equiv \frac{\Delta_{\mathcal{T}}^2}{\Delta^2_{\mathcal{R}}}\,, \label{eq:r1}
\end{align}
where the tensor power spectrum $\Delta^2_{\mathcal{T}}$ in the warm inflation scenario could receive additional contributions from a stimulated emission of gravitons in thermal equilibrium with the radiation bath~\cite{Bhattacharya:2006dm}.\footnote{For other effects that potentially alter the scalar and tensor spectra, see refs. \cite{Namba:2015gja,Peloso:2016gqs}.} Here, we do not include such an effect since it would require the temperature of the thermal bath to be $T \sim M_{\rm pl}$, which is much higher than what is achieved during warm inflation. We then assume that the tensor power spectrum is not affected by the dissipative dynamics and is unchanged with respect to the result for the cold inflation scenario,
\begin{equation}
    \Delta^2_{\mathcal{T}} = \frac{2}{\pi^2} \frac{H^2}{M_{\mathrm{pl}}^2}\,.
\end{equation}
Substituting eq.~\eqref{eq:PS} into eqs.~\eqref{eq:ns1} and~\eqref{eq:r1}, we derive the generalized formulas for the spectral tilt and the tensor-to-scalar ratio:
\begin{align}
    n_s-1&=4\frac{{\rm d}\ln H}{{\rm d}N_e}-2\frac{{\rm d}\ln \dot{\phi}}{{\rm d}N_e}+\left(1+2 n_{\mathrm{BE}}+\frac{2 \sqrt{3} \pi Q}{\sqrt{3+4 \pi Q}} \frac{T}{H}\right)^{-1}\,\left\{ 2n_{\mathrm{BE}}^2e^{\frac{H}{T}}\,\frac{H}{T}\left(\frac{{\rm d}\ln T}{{\rm d}N_e}-\frac{{\rm d}\ln H}{{\rm d}N_e}\right)\right. \nonumber \\
    &\left. +\frac{2 \sqrt{3} \pi Q}{\sqrt{3+4 \pi Q}} \frac{T}{H}\left[ \left(\frac{3+2\pi Q}{3+4\pi Q}\right)\frac{{\rm d}\ln Q}{{\rm d}N_e}+\frac{{\rm d}\ln T}{{\rm d}N_e}-\frac{{\rm d}\ln H}{{\rm d}N_e}\right]\right\} +\frac{ G^\prime(Q)}{G(Q)}Q\frac{{\rm d}\ln Q}{{\rm d}N_e}, \label{eq:ns2}\\
     r&=\frac{16 \epsilon_{w}}{1+Q}\left(1+2 n_{\mathrm{BE}}+\frac{2 \sqrt{3} \pi Q}{\sqrt{3+4 \pi Q}} \frac{T}{H}\right)^{-1} \frac{1}{G(Q)}, \label{eq:r2}
\end{align}
where $G^\prime(Q)\equiv {\rm d}G(Q)/{\rm d}Q$. Finally, in relation to the constraint on the amplitude of the scalar perturbation, a generic slow-rolling inflaton model must satisfy a strict upper bound on the value of the coupling term $\lambda_\phi$ multiplying the quartic self-interaction $\phi^4$ in the inflaton Lagrangian. This was derived at first in ref.~\cite{Adams:1990pn} for standard cold inflation and recently generalized for a strongly dissipative regime to be~\cite{Montefalcone:2022owy}:
\begin{equation}
        \lambda_\phi \lesssim 10^{-9} Q^{-\frac{4b_G}{3}}, \label{eq:lambda_bound}
\end{equation}
where $|b_G|\sim\mathcal{O}(1)$ is the largest power-law exponent of $Q$ in $G(Q)$, which is generally $b_G \geq0$ for $c\geq0$ and $b_G <0$ otherwise. For this, a strictly positive coefficient $c>0$ for the temperature dependence of the dissipation rate leads to a negative correlation between $\lambda_\phi$ and the parameter $Q$, implying that for large values of the dissipation strength $Q$ the constraint in eq.~\eqref{eq:lambda_bound} is significantly tighter than in cold inflation~\cite{Montefalcone:2022owy}.

\section{Warm Natural Inflation \label{sec:3}}

\subsection{General expressions}
\label{sec:3.1}

We consider the following potential for natural inflation:
\begin{equation}
    V(\phi)=\Lambda^4 \Big[ 1+ \cos(\phi/f)\Big], \label{eq:3.1}
\end{equation}
where $\Lambda^2=m_\phi f$ and $m_\phi$ and $f$ are respectively the mass and the decay constant of the axion-like particle which in this scenario plays the role of the inflaton field. Inspired by refs.~\cite{Berghaus:2019whh, Berghaus:2020ekh, DeRocco:2021rzv}, we assume the inflaton couples to a pure Yang-Mills gauge group through the Lagrangian term
\begin{equation}
    \mathcal{L}_{\mathrm{int}} \propto \frac{\phi}{f} \,\mathrm{Tr}\,\mathcal{G}\Tilde{\mathcal{G}}\,,
\end{equation}
where $\mathcal G$ is the gauge field and $\Tilde{\mathcal{G}}$ its dual. We parameterize the dissipation rate as:
\begin{equation}
    \label{eq:tempdepofdissipation}
    \Gamma(T)=\gamma_c \left(\frac{T^c}{f^{c-1}}\right)\,,
\end{equation}
where $\gamma_c$ is a dimensionless factor proportional to the coupling constant between the inflaton field $\phi$ and the gauge field, and $c$ is defined in eq.~\eqref{def:c}.\footnote{Note, that the dimensionless constant $\gamma_c$ could still be dependent on the decay constant $f$, according to the underlying microphysical theory, e.g.\ see eq.~\eqref{eq:gamma3} below.} Motivated by previous model constructions in the literature, we consider the cases $c=\{1,3\}$~\cite{Berghaus:2019whh, Berghaus:2020ekh}.\footnote{See section~\ref{sec:4.2} for more details on these constructions.}

Using eqs.~\eqref{eq:SRp1} and~\eqref{eq:SRp2}, the slow-roll parameters can be written as:
\begin{align}
    \epsilon_{w} &\equiv \frac{\epsilon_V}{1+Q}=\frac{1}{2(1+Q)} \frac{M_{\mathrm{pl}}^{2}}{f^{2}} \frac{\sin ^{2} \phi / f}{(1+\cos \phi / f)^{2}}, \label{eq:SRp11} \\
    \eta_{w} &\equiv \frac{\eta_V}{1+Q}=-\frac{1}{(1+Q)} \frac{M_{\mathrm{pl}}^{2}}{f^{2}} \frac{\cos \phi / f}{1+\cos \phi / f}, \label{eq:SRp22}
\end{align}
with $\beta_{w}=0$ since $\Gamma_{,\phi} \equiv \partial \Gamma/\partial \phi = 0$.

Moreover, we can combine eqs.~\eqref{eq:2.2},~\eqref{eq:SR1},~\eqref{eq:SR3}, and~\eqref{eq:tempdepofdissipation} to derive general relations for $H$, $Q$ and $T$ in terms of $\phi$:
\begin{align}
    H &= \frac{m_\phi f}{M_{\mathrm{pl}}}\sqrt{\frac{1+\cos(\phi/f)}{3}}\,, \\
    T &= \left[ \frac{Q}{(1+Q)^2}\,\frac{1}{4\alpha_1}\, \frac{9 M_{\mathrm{pl}}^6}{f^4 m_\phi^2}\,\frac{\sin^2(\phi/f)}{[1+\cos(\phi/f)]^3}\right]^{1/4}\,, \label{eq:3.5} \\
    Q^{4-c}(1+Q)^{2c} &= \frac{M_{\mathrm{pl}}^{2(2+c)}m_\phi^{2(c-2)}\gamma_c^4}{9 f^{4c} \alpha_1^c}\, \frac{[\sin^2(\phi/f)]^{2c}}{[1+\cos(\phi/f)]^{2+c}}
    \equiv \frac{\xi}{\Tilde{f}^{4c}} \, \frac{[\sin^2(\Tilde{\phi})]^{2c}}{[1+\cos(\Tilde{\phi})]^{2+c}}, \label{eq:3.6}
\end{align}
where in the last step in eq.~\eqref{eq:3.6} we have introduced the notation
\begin{equation}
\Tilde{f}\equiv f/M_{\mathrm{pl}} \,\, , \quad \Tilde{\phi}\equiv\phi/f, \label{eq:3.6.1}
\end{equation}
and we have defined the dimensionless parameter
\begin{align}
    \xi& \equiv \frac{\gamma_c^4}{9 \alpha_1^c}\,\left(\frac{m_{\phi}}{M_{\mathrm{Pl}}}\right)^{2(c-2)}, \label{eq:3.7}
\end{align}
which from now on we will use to characterize $Q$. In other words, for fixed values of $c$, $f$ and $\phi$, $Q$ is uniquely determined by $\xi$ through eq.~\eqref{eq:3.6}.

Inflation ends when one of the slow-roll parameters defined above violates the slow-roll condition, i.e.\ when
\begin{equation}
    \epsilon_V=1+Q, \quad \text{or } \quad |\eta_V|=1+Q. \label{eq:3.8}
\end{equation}
Since both $\epsilon_V$ and $\eta_V$ are simply functions of $\Tilde{\phi}$ and $\Tilde{f}$, it follows that for a given value of $\Tilde{f}$, we can determine the field value at the end of inflation $\Tilde{\phi}_{\mathrm{end}}$ as a function of $\xi$ using the conditions in eq.~\eqref{eq:3.8} combined with eq.~\eqref{eq:3.6}.

\subsection{Existence of a slow-roll regime}
\label{sec:3.1.2}

The existence of a slowly-rolling regime in natural inflation generally depends on the value of the decay constant $f$ and, in the context of warm inflation, on the dissipation strength $Q$. To clarify the connection between the allowed values of $f$ and $Q$, it is useful to first review the standard case of cold natural inflation ($Q=0$). This was done in ref.~\cite{Freese:2004un} and we review the derivation in more detail in Appendix~\ref{appendix:A}.  
The slow-roll condition in standard cold inflation requires  both $\epsilon_V < 1$ and $|\eta_V| < 1$. 
To summarize the results shown in the Appendix, in the cold scenario we obtain two bounds on the value of the decay constant:
(1) To obtain any slowly rolling regime at all requires $\tilde{f}>\sqrt{(\sqrt{2}-1)/2}$; we call this bound a slow-roll existence constraint (ESRC) and (2) a broad slow-roll constraint (BSRC)\footnote{By broad, we mean a large enough regime that easily allow to sustain a slow-rolling field for the number of $e$-folds necessary for a successful inflationary model.} which instead requires $\Tilde{f}>1/\sqrt{2}$. A similar analysis can be conducted in the warm inflation scenario. By making the simplifying assumption of a constant dissipation strength $Q$, we can interpret $\sqrt{1+Q}\Tilde{f}$ as an effective decay constant $\Tilde{f}_w$, and rewrite eqs.~\eqref{eq:SRp11} and~\eqref{eq:SRp22} according to:
\begin{align}
\epsilon_{w}= \frac{1}{2\Tilde{f}_w^2}\frac{\sin ^{2} \Tilde{\phi}}{(1+\cos \Tilde{\phi})^{2}}, \quad
\eta_{w}=-\frac{1}{\Tilde{f}_w^2}  \frac{\cos\Tilde{\phi} }{1+\cos \Tilde{\phi}}.
\end{align}
With this new parameterization, we note that $\epsilon_w$ and $\eta_w$ match $\epsilon_V$ and $\eta_V$ from cold inflation, for $\Tilde{f} \to \Tilde{f}_w$. Thus, we can recover the bounds from cold inflation which now apply to $\Tilde{f}_w$. Therefore, in warm natural inflation
the ESRC and BSRC respectively require $\tilde{f}_w>\sqrt{(\sqrt{2}-1)/2}$ and $\Tilde{f}_w>1/\sqrt{2}$, i.e.
\begin{align}
    \text{ESRC: }\quad \Tilde{f}&>\sqrt{\frac{\sqrt{2}-1}{2(1+Q)}}, \label{eq:WI_SRc1}\\
    \text{BSRC: }\quad \Tilde{f}&>\frac{1}{\sqrt{2(1+Q)}}.  \label{eq:WI_SRc2}
\end{align} 
These can be thought as a generalization of the bounds on the decay constant $\Tilde{f}$ from standard cold inflation. For $Q \ll 1$, we recover the cold inflation result, while for $Q\gg 1$ the bounds on $\tilde{f}$ are strongly alleviated. More generally, for a decay constant $\tilde{f}$ that violates either or both the ESRC and BSRC from standard cold inflation, there is an associated minimum value of $Q$ that restores these bounds in the context of warm inflation. Specifically, by inverting eqs.~\eqref{eq:WI_SRc1} and~\eqref{eq:WI_SRc2} we find:
\begin{eqnarray}
    \text{ESRC:}& \quad \text{for\,} \tilde{f} <(\sqrt{2}-1)/2,\quad &Q>\frac{\sqrt{2}-1}{2\tilde{f}^2}-1, \label{eq:WI_SRc3} \\
    \text{BSRC:}& \text{for\,} \tilde{f} <1/\sqrt{2}, \quad &Q>\frac{1}{2\tilde{f}^2}-1. \label{eq:WI_SRc4}
\end{eqnarray}
For the case $\tilde{f}= 1$, corresponding to the decay constant being equal to the reduced Planck mass, these conditions are satisfied both in cold and warm inflation models. However, the bound on $Q$ rapidly increases as the value of the decay constant is chosen below the Planck scale. For instance, for $\tilde{f}=0.5$, the BSRC bound requires $Q>1$, while for $\tilde{f}=0.1$ the requirement is $Q\gtrsim 50$. In summary, warm natural inflation can be successfully achieved for sub-Planckian values of the decay constant $\Tilde{f}\ll 1$ within a strongly dissipative regime $Q \gg 1$. As shown below, such large values of the dissipation constant have severe effects on CMB observables.

\subsection{Number of $e$-folds}

The number of $e$-folds is defined as:
\begin{align}
    \label{eq:3.9}
    N_{e} \equiv \ln \left(a_{\mathrm{end}} / a_{k}\right)&=\int^{t_{\mathrm{end}}}_{t_{k}} H {\rm d} t,
\end{align}
where $a_k$ is the value of the scale factor when the scale $k$ crosses the horizon ($k=a_k H$), and $a_\mathrm{end}$ is the value of the scale factor when inflation ends. To resolve the flatness problem, a minimum number of $e$-folds is required. For the case in which the Hubble rate remains constant during inflation and reheating is instantaneous, the number of $e$-folds at which observable scales cross the horizon is equal to~\cite{Liddle:2003as}:
\begin{equation}
    N_{e}=68.5 +\frac{1}{4}\ln\left(\frac{V_{\mathrm{hor}}}{M_{\mathrm{pl}}^4}\right) . 
\end{equation}
For the natural inflaton potential considered in eq.~\eqref{eq:3.1} we find $V_{\mathrm{hor}}\simeq \Lambda^4$. Therefore, from the above equation it follows that for $\Lambda\sim\mathcal{O}(10^{14-18})\,\mathrm{GeV}$, we require $N_e\gtrsim 60$. It turns out that for the inflationary models analyzed in this work, the scale of the inflaton potential must be $\Lambda \gtrsim 10^{14}\,\mathrm{GeV}$ in order to satisfy the observational constraints from the CMB. Thus, for the remainder of the paper, we can safely fix $N_e\simeq 60$ and use the integral from eq.~\eqref{eq:3.9} to obtain the field value $\Tilde{\phi}_{\mathrm{CMB}}\equiv \Tilde{\phi}(k_*) $ at horizon
crossing of the pivot scale $k_* = 0.05\,\mathrm{Mpc}^{-1}$ (as a reminder, $\Tilde{\phi} = \phi/f$):
\begin{align}
\text{(eqs.~\eqref{eq:SR1}+\eqref{eq:SR2}):} \quad N_e &=\int_{\phi_{\mathrm{end }}}^{\phi_{\mathrm{CMB}}} \frac{1+Q}{M_{\mathrm{pl}}^{2}} \frac{V}{V_{, \phi}} {\rm d} \phi \\
\text{(eq.~\eqref{eq:3.1}):} \quad &=\Tilde{f}^2\,\int_{\Tilde{\phi}_{\mathrm{CMB}}}^{\Tilde{\phi}_{\mathrm{end}}} (1+Q) \frac{1+\cos \Tilde{\phi}}{\sin \Tilde{\phi}} {\rm d} \Tilde{\phi}.
\end{align}
By additionally fixing $\Tilde{f}$, we can use the previously determined values of $\tilde{\phi}_{\mathrm{end}}(\xi)$ and evaluate the above integral numerically to determine $\Tilde{\phi}_{\mathrm{CMB}}$ as a function of $\xi$.

\subsection{CMB observables}
\label{sec:3.3}

Once we determine the field value at the horizon crossing of the CMB pivotal scale $k_{*}$ and impose the constraint on the overall amplitude of the primordial power spectrum at this scale, we can finally work out the model's predictions such as the spectral index and tensor-to-scalar ratio. As emphasized above, the growth factor $G(Q)$ in eq.~\eqref{eq:PS} can only be determined numerically by solving the full set of perturbation equations for the metric, inflation field and radiation bath~\cite{Graham:2009bf,Bastero-Gil:2011rva,Ramos:2013nsa,Bastero-Gil:2014jsa}. Building on the method of these previous works, we find that, for the natural cosine potential in eq.~\eqref{eq:3.1}, $G(Q)$ takes the following form, respectively for a linear ($c=1$) and cubic ($c=3$) dissipation rate:\footnote{See also Appendix B in ref.~\cite{Das:2020xmh} for more details on the complete set of perturbation equations that include the background evolution, considered in warm inflation.}
\begin{align}
    G_{\mathrm{linear}}\left(Q\right) &\simeq 1+0.189\,Q^{1.642}+0.0028\,Q^{2.729}, \label{eq:G_1}\\
    G_{\mathrm{cubic}}\left(Q\right) &\simeq 1+3.703\,Q^{2.613}+0.0011\,Q^{5.721}. \label{eq:G_3}
\end{align}

Given our numerical fits for $G(Q)$,\footnote{In section~\ref{sec:4.1} we comment briefly on the consistency between our numerical fits for $G(Q)$ and those derived already in the literature for a $\phi^4$ potential~\cite{Bastero-Gil:2016qru, Benetti:2016jhf}.} we now have all the ingredients to evaluate the spectral index and tensor-to-scalar ratio from eqs.~\eqref{eq:ns2} and~\eqref{eq:r2} and compare the obtained results with most recent constraints from observations:
\begin{align}
    n_s & = 0.9649\pm 0.0042,& \quad \text{at 68\% CL,~\cite{Planck:2018jri}} \label{eq:ns_Planck} \\
    r &\lesssim 0.036, & \quad \text{at 95\% CL.~\cite{BICEP:2021xfz}} \label{eq:r_Planck}
\end{align}
Without making any further assumption on the strength of the parameter $Q$, the system of equations that we need to evaluate to obtain $r$ and $n_s$ is very complex and can only be solved numerically. In order to gain an analytical understanding of the effects of the warm inflation scenario on the values of the CMB observables it is useful to study eqs.~\eqref{eq:ns2} and~\eqref{eq:r2} in the strong dissipative regime ($Q \gg 1$). Generally, when thermal effects dominate and for a dissipation rate with a positive temperature dependence $(c>0)$, the radiation will mainly play the role of a source term for the inflaton field fluctuations, inducing an amplification in
the scalar power spectrum.\footnote{In general, one should consider another competing effect to the one above mentioned. In fact, the decay of the inflaton into radiation is effectively playing the role of an additional friction term beyond the usual Hubble friction. This should cause a shift in the point of the potential probed by CMB observations and consequently produce a decrease of $n_s$ and an increase in $r$~\cite{Berera:2018tfc}. For the models presented here, this effect is subdominant and thus can be neglected in a qualitative analysis.} On the other hand, we do not expect  the radiation to affect the tensor fluctuations, which remain unaltered. The overall effect generally induces an increase in $n_s$ and a decrease in $r$ with respect to the cold case~\cite{Berera:2018tfc}. In formulas for $Q \gg 1$ we have:
\begin{align}
    r&\approx \frac{16 \epsilon_V}{a_G\sqrt{3\pi}}\frac{H}{T}\left(\frac{1}{Q}\right)^{\frac{5}{2}+b_G},\label{eq:r_SR} \\
    n_s&\approx1+\frac{1}{Q}\left\{\frac{(2b_G c+4b_G-9)\epsilon_V+[-(3+2b_G)c+6]\eta_V}{4+c}\right\}, \label{eq:ns_SR}
\end{align}
where we parameterized $G(Q)\approx a_G Q^{b_G}$ and used the identities from eq.~\eqref{eq:d1} to~\eqref{eq:d4}. From eq.~\eqref{eq:r_SR}, we can see that $r_{\mathrm{warm}}\sim r_{\mathrm{cold}}/ Q^{\frac{5}{2}+b_G}$ which confirms our intuition for a significantly reduced tensor-to-scalar ratio in the strong dissipative regime. With respect to the spectral index, we can use the functions $G(Q)$ in eqs.~\eqref{eq:G_1}-\eqref{eq:G_3} for the specific cases $c=\{1,3\}$. In the strong dissipative regime, for $Q\gtrsim50$ ($c=1$) and $Q\gtrsim 15$ ($c=3$), the terms with the largest power law dependence on $Q$ dominate in $G(Q)$:  specifically for $c=1$ and $c=3$, the terms with $b_G = 2.729$ and $b_G=5.721$ dominate respectively.\footnote{Results for more general values of $Q$ will be discussed later in section~\ref{sec:4.1}.} Here we find:
\begin{align}
    \text{$\{c,b_G\}=\{1,2.729\}$: } \quad n_{s,1}&\approx 1+ \frac{7.37\epsilon_V-2.46\eta_V}{5Q}> 1, \label{eq:ns.SR1}\\
    \text{$\{c,b_G\}=\{3,5.721\}$: } \quad n_{s,3}&\approx 1+ \frac{48.21\epsilon_V-37.33\eta_V}{7Q}> 1, \label{eq:ns.SR3}
\end{align}
since $\epsilon_V>\eta_V$ $\forall \phi\in(0,\pi)$. Therefore, in the strong dissipative regime, the spectral tilt is blue-shifted $(n_s>1)$ which clearly violates the observational constraints from the CMB in eq.~\eqref{eq:ns_Planck} that instead predict a slightly redshifted spectrum. This sets an important constraint on the maximum allowed value of $Q$ that these models can support in order to match observations. As explained in section~\ref{sec:3.1.2}, this in turn translates into a lower bound for the axion decay constant $f$. 

\section{Results \label{sec:results}}

In this section we present our results for the case in which the dissipation rate in eq.~\eqref{eq:tempdepofdissipation} increases with temperature as $T^c$ with $c=\{1,3\}$.  As discussed further in section~\ref{sec:4.1}, as we take the value of the decay constant $f$ to decrease, the allowed range of $Q$ values becomes ever narrower, so that for a given $c>0$, there is a minimum allowed value of $f$ not far below the Planck scale. Figure~\ref{fig:2} shows our results for the quantities $r$ and $n_s$, for given values of $f$ and as a function of $Q$. Our main results can be seen in figure~\ref{fig:1}, which compares the predictions of models in the $r$-$n_s$ plane with CMB and Baryon Acoustic Oscillation (BAO) data. 

For all of the models discussed below, we checked numerically that they satisfy the bound on the quartic interaction term in eq.~\eqref{eq:lambda_bound} which in warm natural inflation is $\lambda_\phi=m_\phi^2/(24f^2)$.

\subsection{Comparison with data for
linear and cubic dissipation rates \label{sec:4.1}}

\begin{figure}[ht!]
    \centering
    \includegraphics[scale=1]{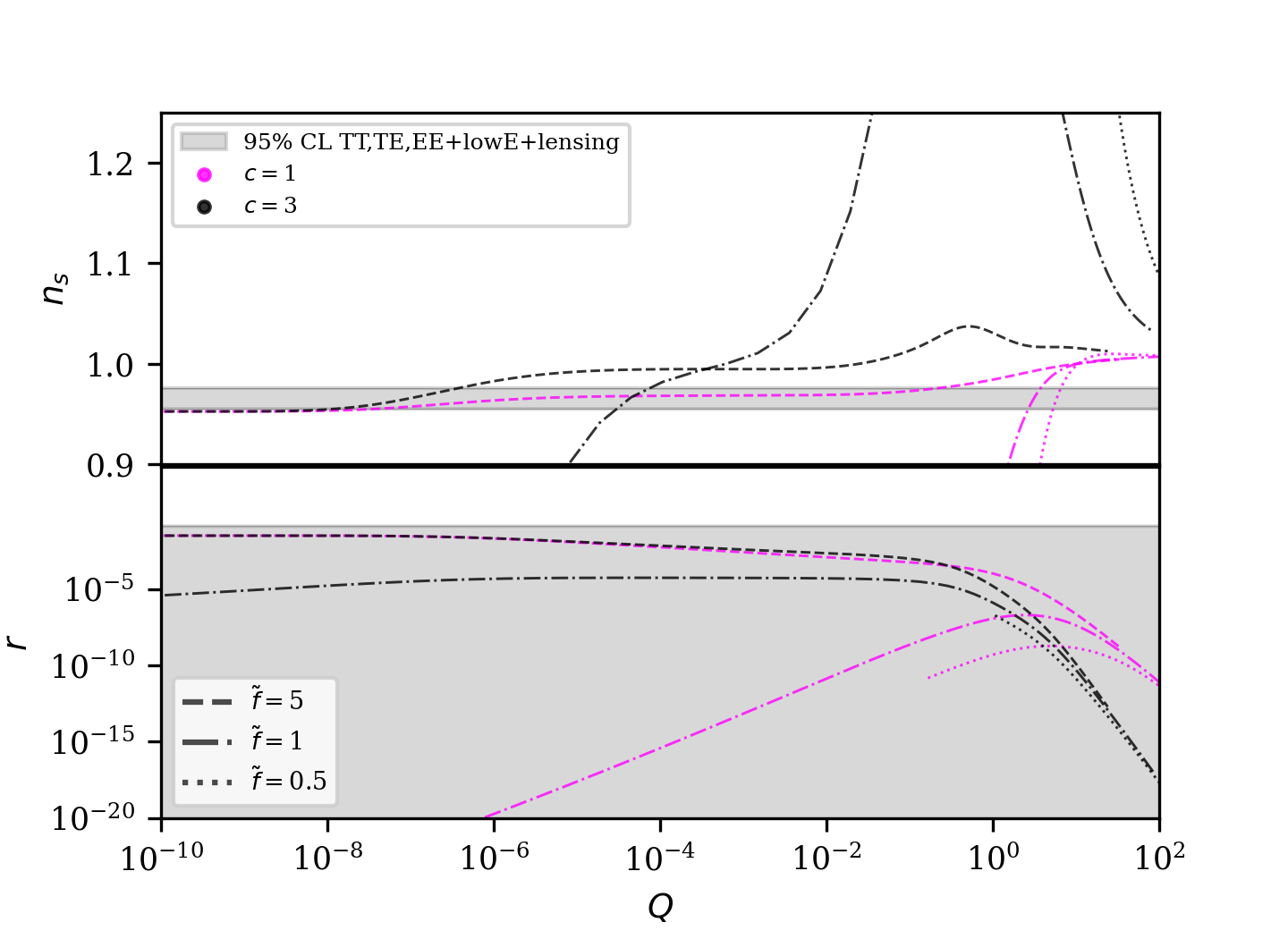}
    \caption{Scalar spectral index $n_s$ (upper panel) and tensor-to-scalar ratio $r$ (lower panel) as a function of the dissipation parameter $Q$ for $c=\{1,3\}$ and $\tilde f = f/M_{\mathrm{pl}} =  \{5,1,0.5\}$ as shown in the legend of the lower panel. The values of $n_s$ and $r$ consistent with \textit{Planck} 2018 data at 95\% CL are shown in both panels as the horizontal light gray region. Upper panel: The value of $n_s$ generally rises with increasing $Q$. Further, the maximum value of $Q$  for which $n_s$ is consistent with \textit{Planck} 2018 data increases for smaller values of $\tilde f$. Indeed there is no acceptable $Q$ value for the case of $c=3$ with $\tilde f = 0.5$ (although not shown the curve $n_s$ vs $Q$ for the $\tilde{f}=0.5$ case plateaus at a value $>1$ as $Q$ increases). Lower panel: For smaller values of $\tilde f$, the value of $r$ decreases. For $\tilde{f}=0.5$, the lowest values of $Q$ shown are due to the lower limit arising from the requirement of the existence of a broad slowly-rolling regime.}
    \label{fig:2}
\end{figure}

\begin{figure}[ht!]
    \centering
    \includegraphics[scale=1]{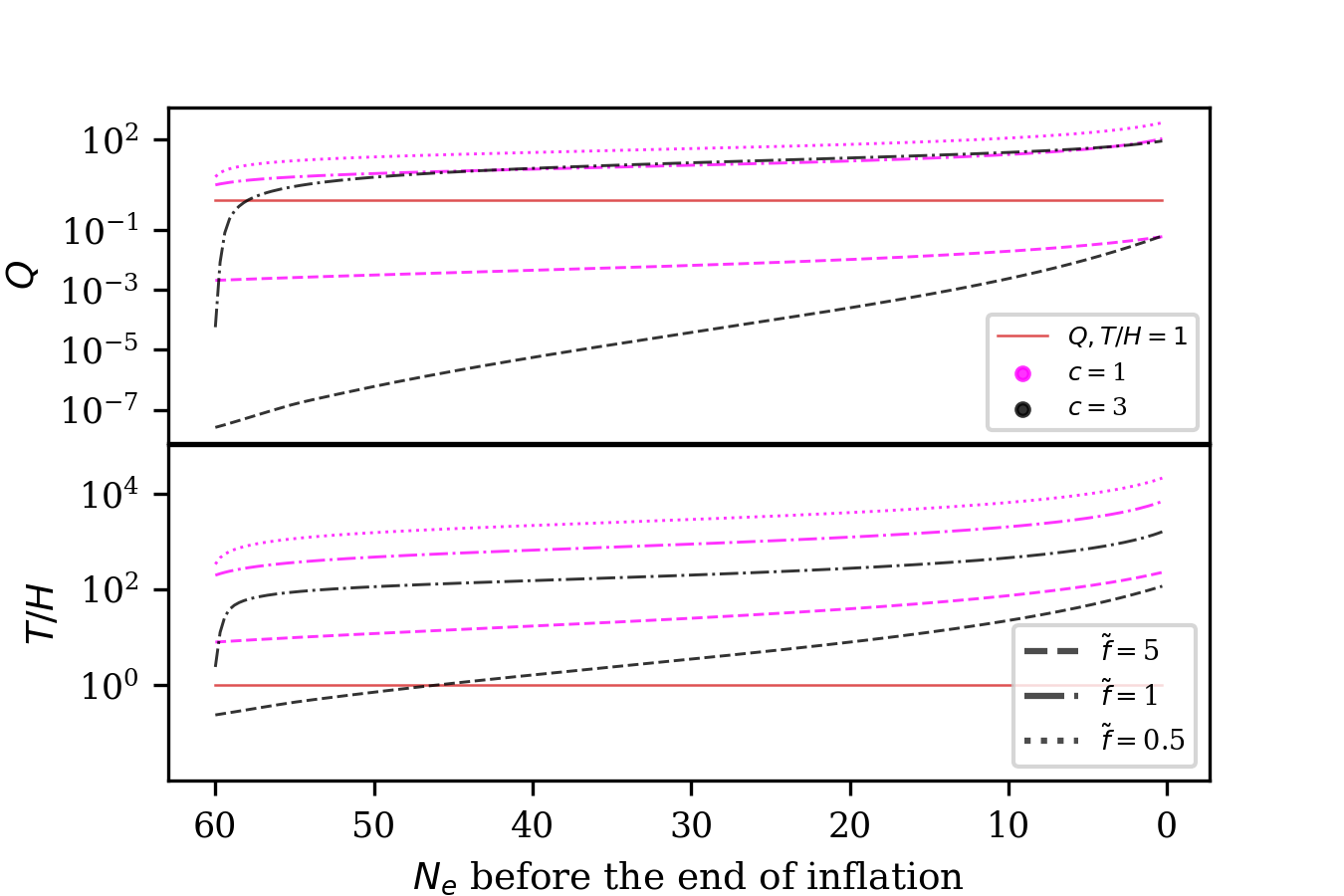}
    \caption{The evolution of the dissipation strength $Q$ (upper panel) and the $T/H$ ratio (lower panel) for $c=\{1,3\}$ and $\tilde f = f/M_{\mathrm{pl}} =\{5,1,0.5\}$ as shown in the legend of lower panel. The horizontal axis is the number of $e$-folds before the end of inflation, so the evolution starts from the left at $N_e = 60$ and evolves to the right to $N_e=0$. All the curves are obtained by fixing a specific value of $\xi$ that we know produces a inflationary model within the observational constraints on $r$ and $n_s$.}
    \label{fig:3}
\end{figure}

We use the methodology described in the previous sections to analyze warm natural inflation for both a linear and cubic temperature-dependent dissipation rate. We fix a priori the number of $e$-folds $N_e$ and the decay constant $f$. Specifically we take $N_e=60$ and $\Tilde{f}=\{5,1,0.5\}$.\footnote{We checked the numerical results obtained for $N_e=\{55,65\}$ and found no significant difference from the $N_e=60$ case presented here.} This allows us to express all the quantities of interest in terms of a single parameter $\xi$ defined in eq.~\eqref{eq:3.7}, which qualitatively characterizes the magnitude of the dissipative effect. For each model, we map the region of the parameter space in the $r$-$n_s$ plane consistent with the observational constraints. This region translates into bounds on the allowed values of $\xi$ which in turn can be interpreted as constraints on the dissipation strength $Q$ at horizon crossing of the CMB pivotal scale $k_{*}$, and on the dimensionless factor of proportionality of the dissipation rate $\gamma_c$. These results and the corresponding constraints are summarized in figure~\ref{fig:2} through figure~\ref{fig:3} and in table~\ref{table:1}, where $Q_*$ is evaluated at horizon crossing of the CMB pivot scale.

\begin{table}[ht!]
\def\arraystretch{1.5}
\centering
    \begin{tabular}{l>{\raggedleft}p{0.33\textwidth}>{\raggedleft\arraybackslash}p{0.3\textwidth}>{\raggedleft\arraybackslash}p{0.3\textwidth}}
    \toprule
     \hline
     \multicolumn{3}{c}{Observational constraints on $\gamma_c$ and $Q_{*}$}\\
     \hline
Model\,($c,\Tilde{f}$) & Allowed range of $Q_{*}$ & Allowed range of $\gamma_c$\\
    \hline
    $(3,5)$ & $[2.06\times 10^{-8},2.82\times 10^{-7}]$ & $[1.53\times10^5,3.04\times10^5]$ \\
    $(3,1)$ & $[3.03\times10^{-5},6.31\times 10^{-5}]$ & $[7.65\times10^6,7.71\times10^6]$\\
    $(1,5)$ & $[8.19\times 10^{-8}, 0.14]$ & $[5.46\times 10^{-7},8.3\times 10^{-3}]$\\
    $(1,1)$ & $[2.89, 3.77]$ & $[3.6\times 10^{-2},3.8\times 10^{-2}]$ \\
    $(1,0.5)$ & $[5.77,7.19]$ & $[4.01\times 10^{-2},4.05\times 10^{-2}]$\\
    \hline
    \bottomrule
    \end{tabular}
    \caption{Range of values of $Q$ and $\gamma_c$ allowed by CMB data (more precisely, the 95\% CL from the \textit{Planck} 2018 results) for dissipation scaling as $T^c$ with $c=\{1,3\}$. Here, $Q_*$ is evaluated at horizon crossing of the CMB pivot scale. The parameters $c$ and $\gamma_c$ are the exponent and normalization of the temperature dependence of the dissipation rate in eq.~\eqref{eq:tempdepofdissipation}, respectively. As $\tilde f$ decreases, the allowed range of $Q_*$ narrows, so that there is no acceptable value of $Q$ left for the case of $c=3$ with $\tilde f = 0.5$.}
    \label{table:1}
\end{table}

For both $c=\{1,3\}$ and $f\geq M_\mathrm{pl}$, a region of the parameter space consistent with the observational constraints on $r$ and $n_s$ at the 1$\sigma$ level exists. Moreover, in contrast with the cubic scenario, the linear case also predicts a small but non-zero parameter space for a marginally sub-Planckian decay constant ($\Tilde{f}=0.5$). Overall, the allowed parameter space always lies in the weak or moderate dissipative regime in agreement with the analysis presented in section~\ref{sec:3.3}. In fact, as shown in figure~\ref{fig:2}, for $Q\gg 1$ the spectral index is blue-shifted in all cases. Figures~\ref{fig:2} and~\ref{fig:1} also illustrate four important aspects of this class of models, which we describe in detail below:

\begin{figure}[ht!]
    \centering
    \includegraphics[width=\linewidth]{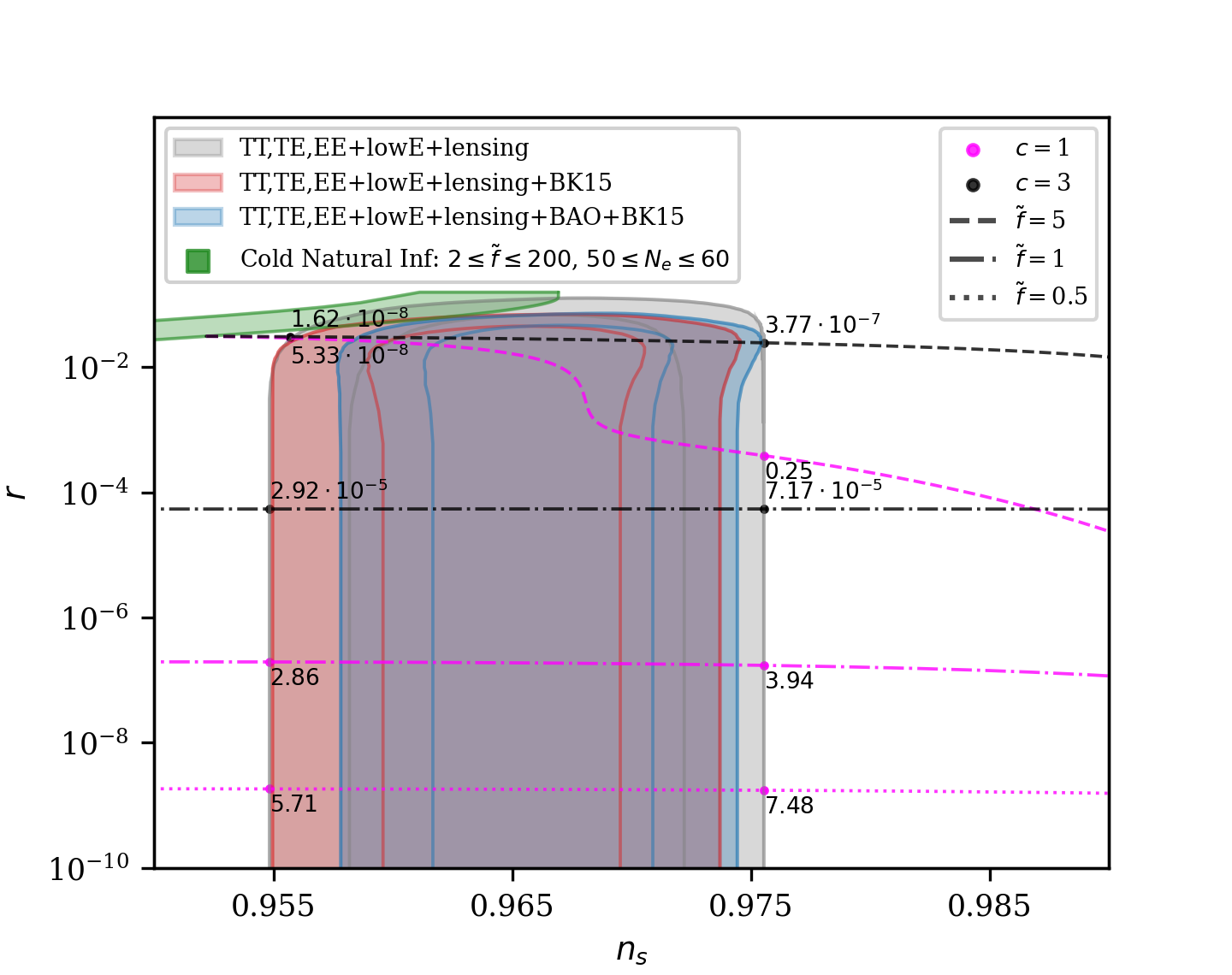}
    \caption{The marginalized joint 68 and 95\% CL regions for the spectral index $n_s$, and the tensor-to-scalar ratio $r$, obtained from \textit{Planck} 2018 and lensing data alone, and their combinations with BICEP2/Keck Array (BK15) and BAO data, confronted with the predictions of cold (green) and warm natural inflationary models with $c= \{1,3\}$ (where $c$ is the power law characterizing the temperature dependence of the dissipation). The curves are obtained by varying $Q$ for fixed values of the decay constant $\tilde f = f/M_{\mathrm{pl}}$ as marked in the legend.  The numbers at the intersections between these curves and the 95\% CL region from \textit{Planck} 2018 show the corresponding values of $Q$. Overall, as we lower $\tilde f$, the tensor-to-scalar ratio $r$ gets smaller while the allowed values of $Q$ shrinks in range and gets higher in magnitude.}
    \label{fig:1}
\end{figure}

\begin{enumerate}

    \item The tensor-to-scalar ratio $r$ is within the observational constraints at the $2\sigma$ level for all values of $Q$ and decreases rapidly for $Q\gtrsim 1$. This is in agreement with eq.~\eqref{eq:r_SR} which emphasizes the negative power law dependence of $r$ on $Q$ when thermal effects dominate.
    
    \item As $f$ decreases, the region where the spectral index $n_s$ is within the observational constraints moves to higher values of $Q$ and shrinks in size. This is highlighted in table~\ref{table:1} which shows that models with smaller values of $f$ are more tightly constrained. In addition, this feature also clarifies the competing effects that prevent us from obtaining a valid model for a significantly sub-Planckian $f$. In short, to lower $f$ we need a larger value of $Q$ which in turn increases $n_s$. Thus there is only so much room to lower $f$, without requiring $Q \gg 1$ and consequently forcing the spectrum to be blue-shifted. For $\tilde{f}=0.5$, the smallest values of $Q$ shown represent the lower limit arising from the requirement of the existence of a broad slowly-rolling regime and correspond to $Q_{*}\gtrsim 0.2$ and $Q_{*}\gtrsim 1$ for $c=1$ and $c=3$, respectively. These values only roughly match what we derived in eq.~\eqref{eq:WI_SRc4}, which is expected since the dissipation strength is not constant (as we assumed) but clearly evolves during the inflationary period, see figure~\ref{fig:3}, and does so differently for the linear and cubic case.
    
    \item For a given value of $f$, the spectral index $n_s$ becomes blue-shifted at smaller values of $Q$ for $c=3$ compared to $c=1$. This is a consequence of the sharper temperature dependence of the cubic dissipation rate which in turn causes a larger enhancement of the scalar power spectrum, compared to the linear case. In formulas, this is emphasized by the last term in eq.~\eqref{eq:ns2} ($\propto Q G^\prime (Q)/G(Q)$) which dominates in the strong regime and is always larger for $c=3$ compared to $c=1$. This feature also explains why a linear dissipation rate is able to accommodate a slightly smaller decay constant $f$ than the cubic case. We know from section~\ref{sec:3.1.2} that the minimum initial value of $Q$ needed to sustain $\sim 60$ $e$-folds of inflation in the slow-roll regime increases for lowering $\Tilde{f}$. For $c=3$ and $\Tilde{f}=0.5$, this value is already large enough to make the spectrum blue-shifted. On the other hand, for $c=1$ and $\Tilde{f}=0.5$, there is still a small but non-zero range where the spectral index is within the observational constraints. We have checked numerically for the lowest allowed value of $f$ that is compatible with the observational constraints from the CMB. We found that for $c=1$: $\tilde{f}_{\mathrm{min}}=0.3$, and for $c=3$: $\tilde{f}_{\mathrm{min}}=0.8$, respectively.
    
    \item For $\tilde{f}=5$, we can clearly see in figure~\ref{fig:1} that both cases $c=\{1,3\}$ reduce to the cold natural inflation result. This occurs precisely when $Q_{*}\leq 9.5\times 10^{-11}$ and $Q_{*}\leq 3.7\times 10^{-10}$ respectively for $c=1$ and $c=3$. In fact, we have checked numerically that for all such values of $Q_{*}$, $H>T$ throughout the whole inflationary period and therefore the thermal bath has no influence on the inflaton dynamics.

\end{enumerate}

Our numerical fits for $G(Q)$ shown in eqs.~\eqref{eq:G_1}-\eqref{eq:G_3} approximately match those derived in~\cite{Bastero-Gil:2016qru, Benetti:2016jhf} for a quartic potential. To see this, we have checked numerically that using the fits for $G(Q)$ from refs.~\cite{Bastero-Gil:2016qru, Benetti:2016jhf} (see e.g.\ refs.~\cite{Reyimuaji:2020bkm,Correa:2022ngq}) instead of those derived here in the context of warm natural inflation, has a negligible impact on the observational constraints obtained for these models. As expected, this has to do mainly with the fact that CMB observations, as clearly shown in figures~\ref{fig:2} and~\ref{fig:1}, prefer a weak or moderate dissipative regime, for which we find that the two fits differ at most by $20\%$ $(30\%)$ for $c=1$ $(c=3)$.

To further show the dynamics during inflation, in figure~\ref{fig:3} we follow the prescription in ref.~\cite{Reyimuaji:2020bkm} and plot for each model the evolution of $T/H$ and $Q$ as a function of the number of $e$-folds left until the end of inflation.\footnote{To do so, we choose $N_e=60$ for simplicity and fix a specific value of $\xi$ that corresponds to an inflationary model lying within the observational constraints for $r$ and $n_s$.} As expected we see that both $Q$ and $T/H$ increase during inflation. For $c=3$, $\Tilde{f}=5$, inflation starts in a cold scenario ($T/H<1$) and evolves in the warm scenario ($T/H>1$) via the coupling to the radiation. For all the other models presented here, inflation starts already in a warm context and simply gets warmer during the exponential expansion. With respect to the strength of the dissipation rate $Q$, we see that for $\Tilde{f}=5$, we have $Q<1$ during all the inflation period for both $c=\{1,3\}$. On the other hand for $\Tilde{f}=\{1,0.5\}$, inflation only starts with a $Q\sim \mathcal{O}(1)$ which quickly increases to values $Q>1$ through most of the inflationary period.

In summary, for the cases of linear and cubic friction terms, i.e.\ $c=\{1,3\}$, while we are able to rescue natural inflation in the warm inflation scenario for much smaller values of $f$ compared to the corresponding cold inflation case, we are still unable to decrease $f$ substantially below Planckian scales.

\subsection{Comparison of our results with theoretically motivated axion-like interaction terms
}\label{sec:4.2}

As briefly mentioned in section~\ref{sec:3.1}, our study of a linear and cubic friction term is inspired by theoretically motivated axion-like interaction terms in the literature. Here we compare the constraints derived in this paper with these theoretical models.
In the generic case of a gauge sector (with trace normalization $T_R$ and dimension $d_R$ of the representation) and an associated light fermion with mass $m_f$, the friction term $\Gamma (T)$ was computed to be~\cite{Berghaus:2020ekh}:
\begin{align}
    \Gamma (T)&=\frac{\Gamma_{\mathrm{sph}}}{2 T f^{2}}\left(1 + \frac{24 T_{R}^{2}}{d_{R} T^3} \frac{\Gamma_{\mathrm{sph}}}{\Gamma_{\mathrm{ch}}}\right)^{-1}\,, \label{eq:gamma1} \\
    \Gamma_{\mathrm{sph}}&\equiv\Tilde{\kappa}(\alpha,N_c,N_f)\alpha^5 T^4\,, \\
    \Gamma_{\mathrm{ch}}&\equiv \frac{\kappa\, N_c\,\alpha\, m_f^2}{T}\,,
\end{align}
where $\Gamma_{\mathrm{sph}}$ and $\Gamma_{\mathrm{ch}}$ are the sphaleron transition rate~\cite{Arnold:1996dy} and chirality-violating rate~\cite{Boyarsky:2020cyk}, respectively. Additionally, $\alpha=\Tilde{g}^2/(4\pi)$ represents the coupling; the coefficient $\kappa$ is $\mathcal{O}(1)$, which we set to $\kappa = 1$ henceforth; $\Tilde{\kappa}$ is a $\mathcal{O}(100)$ number which has a weak logarithmic dependence on $\alpha$ and whose exact value depends on the number of colors $N_c$ and flavors $N_f$ of the group. In this context, $\Gamma(T)$ accounts for the frictional effect of sphaleron transitions between gauge vacua. The role of light fermions is to allow chirality-violating processes that diminish the friction associated with sphaleron transitions~\cite{Berghaus:2020ekh}. We will consider two cases of infinite and finite fermion masses. In the limit $m_f \rightarrow \infty$ (case 1), $\Gamma$ reduces to the previously derived form for a pure-gauge sector with a cubic temperature dependence~\cite{Berghaus:2019whh}. For a finite fermion mass $m_f\lesssim (N_c^2\alpha^2)T$ (case 2), the pure-gauge frictional term is suppressed by $\propto (N_c\alpha)^{-4}(m_f/T)^2$ and the effective friction term becomes linear in temperature. In formulas, we get:
\begin{align}
    \text{(case 1):} \quad \Gamma(T)&\simeq \Big(\frac{\Tilde{\kappa} \alpha^5}{2f^2}\Big)T^3; \label{eq:gamma2} \\
    \text{(case 2):} \quad \Gamma(T)&\simeq \Big(\frac{d_R N_c \alpha m_f^2}{48 f^2 T_R^2}\Big)T; \label{eq:gamma3}
\end{align}
which clearly produce a dissipation rate linear and cubic in the temperature, as those analyzed in this work. The estimation of the dissipation rate from the axion-like interaction as in eq.~\eqref{eq:gamma1} is only known to be valid for $m_{\phi} < \alpha^2 T$ and $H < \alpha^2 T$. In addition, $\alpha$ is also bounded from perturbativity and the inflaton thermalization which respectively require $\alpha\lesssim 0.1$~\cite{Moore:2010jd} and $\alpha < 10^{-2}\sqrt{Q}$~\cite{Berghaus:2019whh}. The constraint on the inflaton thermalization simply amounts to imposing the gauge boson-axion scattering rate $\Gamma_{\phi g}\approx \alpha^3 T^3/(32\pi^2 f^2)$~\cite{Masso:2002np,Graf:2010tv} to be much bigger than the Hubble parameter $H$. This bound is also valid in the case of a finite fermion mass (case 2) as the boson-axion scattering rate is always dominant over the fermion-axion interaction~\cite{Berghaus:2020ekh}.
 
It is then of great interest to compare these theoretical bounds on the value of $\alpha$ with the observational constraints on $\gamma_c$ and $Q_{*}$ derived in this work. By combining eqs.~\eqref{eq:gamma2} and~\eqref{eq:gamma3} with eq.~\eqref{eq:tempdepofdissipation}, we obtain the following relationships:
\begin{align}
    \text{(case 1):} \quad \gamma_3=\frac{\Tilde{\kappa} \alpha^5}{2}&\sim \mathcal{O}(10^2)\cdot \alpha^5, \\
    \Rightarrow \alpha &\sim \Big(\frac{ \gamma_3}{10^2}\Big)^{\frac{1}{5}}; \\
    \text{(case 2):} \quad \gamma_1=\frac{d_R N_c \alpha m_f^2}{48 f^2 T_R^2} &\lesssim \frac{d_R N_c^5 \alpha^5 T^2}{48 f^2 T_R^2} \sim \mathcal{O}(1)\cdot \frac{\alpha^5 T^2}{f^2}, \\
    \Rightarrow \alpha &\gtrsim \Big(\frac{f^2 \gamma_1}{T^2}\Big)^{\frac{1}{5}}. \label{eq:alpha_2} 
\end{align}
Substituting the allowed range of values for $\gamma_3$ from table~\ref{table:1} in the expression above gives $\alpha \gtrsim 1$ for both $\Tilde{f}=\{5,1\}$, which violates the constraint on perturbativity. Additionally, the constraint from the inflaton thermalization is in even greater tension since for models with $c=3$ we have $Q_{*}<10^{-5}$, which in turn requires $\alpha<10^{-7}$. In a similar fashion, we compute the observational bounds on $\alpha$ for the linear case and we find $\alpha \gtrsim 50$ which again violates the theoretical constraints on both perturbativity and thermalization, for which setting $Q_{*}\sim \mathcal{O}(1)$ requires $\alpha<10^{-2}$. To summarize, the entirety of the viable parameter space that we obtained in this work strongly violates the theoretical bounds on the cubic and linear axion-like interaction terms. This tension can be alleviated via two main mechanisms. First of all, the limit $\alpha\lesssim 0.1$ set by current lattice calculations is more of a technical reason than a physical one, see ref.~\cite{Moore:2010jd} for more details. Physically, values of $\alpha \sim 1$ are allowed and would ease this bound. Secondly, the inflaton does not need to be thermalized at the beginning of inflation, in which case $\alpha$ would only be bounded from perturbativity. This option is discussed in detail in ref.~\cite{Reyimuaji:2020bkm} for a cubic interaction term and it is shown to be an effective mechanism to alleviate the tension on $\alpha$. 

That being said, it is important to also emphasize that our results are not confined to a specific microphysical origin of the dissipation rate. As long as the dissipation rate has a linear or cubic temperature dependence, our results are valid and thus provide useful constraints for future model constructions of dissipative rates of this kind.

\section{Summary and future outlook \label{sec:5}}

In this paper we studied warm natural inflation. More specifically, we considered the original cosine potential suggested in ref.~\cite{Freese:1990rb}, in the presence of a radiation bath produced as inflation progresses. Extending over previous work in the literature~\cite{Visinelli:2011jy, Reyimuaji:2020bkm, Correa:2022ngq}, we considered warm natural inflation with a linear and cubic temperature-dependent dissipation rate, which represent physically well-motivated cases~\cite{Berghaus:2019whh, Berghaus:2020ekh, DeRocco:2021rzv}. We numerically solve for the evolution of the set of background equations and their perturbations in the case of a cosine potential for the inflaton, and we derive the power spectrum of scalar perturbations as a function of the dissipative rate in warm inflation. This is encoded in the numerical fits in section~\ref{sec:3.3}. For each of these models we mapped their location in the $r$-$n_s$ plane and compared them to CMB data, see figure~\ref{fig:1}.

We found that, in contrast with the standard cold inflation scenario, for $f\geq 1 M_\mathrm{pl}$ warm natural inflation is consistent with observational constraints on $r$ and $n_s$ at the $1\sigma$ level, respectively in a weak dissipative regime for $c=3$ and in a weak or moderate dissipative regime for $c=1$.\footnote{As a reminder to the reader, the weak and moderate dissipative regimes are respectively defined as $Q\lesssim1$ and $1\leq Q\leq 10$ where $Q \equiv \Gamma/3H$.} Indeed, for the $c=3$ case, the requirement is in the regime $Q\ll 1$ (see table~\ref{table:1}), yet the dissipation rate can play an important role during inflation while $T>H$. 

A goal of our study was also to determine whether or not the requirement of a trans-Planckian decay constant ($f\gtrsim M_{\mathrm{pl}})$, required by CMB observations for the case of cold natural inflation but undesirable from the perspective of model-building, could be avoided in the presence of the radiation bath. Intuitively, one would expect sub-Planckian values of $f$ to be easily accommodated as the radiation bath produced while the inflaton is rolling down its potential is effectively playing the role of an additional friction term. This reduces the required field excursion $\Delta \phi$ compared to the cold inflation case for the same number of $e$-folds of inflation, i.e.\ we can accommodate a smaller width of the potential $f$ while still obtaining sufficient inflation. While this conclusion was found to be true for warm natural inflation with a constant dissipation rate~\cite{Visinelli:2011jy}, here we found that there are two competing effects that prevent us from obtaining a successful inflationary model with a sub-Planckian value of $f$. On one hand, it is true that friction allows for a smaller width of the potential for sufficient inflation. However, we also find the following: as $f$ is lowered, the dissipation strength $Q$ must increase in order to maintain the existence of a (broad) slowly-rolling phase; however, a larger $Q$ leads to a larger scalar spectral index $n_s$. Unfortunately, we found that $f$ cannot become significantly sub-Planckian without resulting in $n_s>1$, which is ruled out by CMB data. Specifically, we found that the minimum value of the decay constant that satisfies the cosmological constraints in the $r$-$n_s$ plane shown in figure~\ref{fig:2} is $f_{\rm min} = 0.8\,M_{\mathrm{pl}}$ for $c=3$, and $f_{\rm min} = 0.3\,M_{\mathrm{pl}}$ for $c=1$. 

As a whole, the region of the parameter space consistent with the CMB observational constraints is in tension with the thermal field theory requirements of the cubic and linear axion-like interaction terms that motivated our study in the first place. Nevertheless, our results are not confined to a specific microphysical origin of the dissipation rate and thus provide useful constraints for future model constructions of dissipative rates of this kind.

Finally, an interesting and more speculative alternative is to consider a dissipation rate with an inverse temperature dependence, i.e.\ $c=-1$. In fact, in contrast with the $c>0$ cases, for $c<0$ the scalar power spectrum is suppressed at large values of $Q$~\cite{Bastero-Gil:2011rva}, which in general results in a redder spectrum. Following eq.~\eqref{eq:ns_SR}, we find that for $c=-1$ and $Q\gg 1$ we get:
\begin{align}
    n_{s,-1}&\approx 1+ \frac{(2b_G+9)\eta_V -(9-2b_G)\epsilon_V}{3Q}, \label{eq:ns_c-1}
\end{align}
which is in fact always $<1$ as long as $b_G>-\frac{9}{2}$.\footnote{If we take the growth factor $G(Q)$ derived in ref.~\cite{Motaharfar:2018zyb} we have $b_G=-1.41>-\frac{9}{2}$.} This motivates the possibility to study the $c=-1$ case in the strong regime of warm inflation and for a significantly sub-Planckian value of $f$. This however will require an extension of the range of validity of the growth factor $G(Q)$ for the $c=-1$ case previously derived in ref.~\cite{Motaharfar:2018zyb} to larger values of $Q$, i.e.\ $Q\gtrsim 10^{6}$ for $f=10^{-3}M_\mathrm{pl}$, see eqs.~\eqref{eq:WI_SRc3} and~\eqref{eq:WI_SRc4}, and we therefore leave it to future work.

In summary, other than for the case of constant dissipation throughout the inflationary epoch, it is difficult to obtain a significantly sub-Planckian decay constant even in the case of warm natural inflation.

\section*{Acknowledgements}
We would like to thank Rudnei O. Ramos, Kim V. Berghaus, Will Kinney and Xinyi Zhang for helpful discussions. K.F.\ is Jeff \& Gail Kodosky Endowed Chair in Physics at the University of Texas at Austin, and K.F.\ and G.M.\ are grateful for support via this Chair. K.F.\ and G.M.\ acknowledge support by the U.S.\ Department of Energy, Office of Science, Office of High Energy Physics program under Award Number DE-SC-0022021 as well as support from the Swedish Research Council (Contract No.~638-2013-8993). V.A.\ acknowledges support from the National Science Foundation under grant number PHY-1914679. 

\newpage
\appendix

\section{Existence of a slow-roll regime in standard cold natural inflation \label{appendix:A}}

The slow-roll condition in standard cold inflation requires $\epsilon_V$ and $|\eta_V|$ to be both $<1$. In standard cold natural inflation, characterized by the cosine potential in eq.~\eqref{eq:3.1}, these conditions read:
\begin{align}
\text{($\epsilon_{V}<1$): }& \quad  \frac{1-t^2}{(1+t)^{2}}<2\tilde{f}^2, \label{eq:A.1} \\
\text{($|\eta_{V}|<1$): }& \quad \Big|\frac{t}{1+t}\Big|<\tilde{f}^2, \label{eq:A.2}
\end{align}
where $t=\cos\tilde{\phi}$, and $\tilde{\phi}\equiv \phi/f\in(0,\pi)$ and $\tilde{f}\equiv f/M_{\mathrm{pl}}$. The solution of eq.~\eqref{eq:A.1} is simple and always sets an upper bound on the maximum allowed value of $\tilde{\phi}$:
\begin{equation}
 \text{($\epsilon_{V}<1$): } \quad   \tilde{\phi}<\arccos{\left(\frac{1-2\tilde{f}^2}{1+\tilde{f}^2}\right)}\equiv \tilde{\phi}_\epsilon.
\end{equation}
On the other hand, the solution to eq.~\eqref{eq:A.2} depends on the value of $\tilde{f}$. If $\tilde{f}\geq 1/\sqrt{2}$, the $\eta_V$ constraint sets an upper bound on the maximum allowed value of $\tilde{\phi}$, otherwise for $\tilde{f}< 1/\sqrt{2}$ the constraint imposes a lower bound on $\tilde{\phi}$. In formulas we have:
\begin{equation}
    \text{($|\eta_{V}|<1$): } \begin{cases}
  \tilde{\phi}<\arccos{\left(\frac{-\tilde{f}^2}{1+\tilde{f}^2}\right)}\equiv \tilde{\phi}_{\eta,1} & \text{for } \tilde{f}\geq\frac{1}{\sqrt{2}}, \\   
  \\
  \tilde{\phi}>\arccos{\left(\frac{\tilde{f}^2}{1+\tilde{f}^2}\right)}\equiv \tilde{\phi}_{\eta,2} & \text{otherwise.}    
\end{cases}
\end{equation}
Since $\tilde{\phi}_\epsilon<\tilde{\phi}_{\eta,1}$ for all $\tilde{f}\geq 1/\sqrt{2}$, we can therefore conclude that for such values of the decay constant we have a broad parameter space for the slow-rolling regime, i.e.\ for all $\phi\in(0,\tilde{\phi}_\epsilon)$. On the other hand, for $\tilde{f}<1/\sqrt{2}$, the range of validity of the slow-roll regime shrinks in size and amounts to $\phi\in(\tilde{\phi}_{\eta,2},\tilde{\phi}_\epsilon)$. Therefore, to guarantee a broad slow-roll regime we require $\tilde{f}>1/\sqrt{2}$. This bound is what we call in this work the broad slow-roll constraint (BSRC).

Additionally, as we continue to lower $\tilde{f}$, the constraints on $\epsilon_V$ and $|\eta_V|$ keep pushing on opposite directions, with the value of $\tilde{\phi}_{\eta,2}$ that increases while that of $\tilde{\phi}_{\epsilon}$ decreases. For $\tilde{f}=\sqrt{(\sqrt{2}-1)/2}$ the overlap between the two constraints becomes null and the slow-rolling condition is violated for all $\tilde{\phi}\in(0,\pi)$. This means that to guarantee the existence of a non-zero slow-roll parameter space, we need $\tilde{f}>\sqrt{(\sqrt{2}-1)/2}$. This bound is what we call in this work the existence slow-roll constraint (ESRC). 

\bibliographystyle{JHEP}
\bibliography{bibl.bib}

\end{document}